\pdfoutput=1
\documentclass[lettersize,journal]{IEEEtran}
\usepackage{amsmath,amsfonts}
\usepackage{algorithmic}
\usepackage{algorithm}
\usepackage{array}
\usepackage[caption=false,font=normalsize,labelfont=sf,textfont=sf]{subfig}
\usepackage{textcomp}
\usepackage{stfloats}
\usepackage{url}
\usepackage{verbatim}
\usepackage{graphicx}
\usepackage{cite}
\usepackage{color}
\usepackage{comment}
\usepackage{subcaption}
\usepackage{multirow}
\usepackage[x11names,table]{xcolor}
\usepackage{amssymb}
\usepackage[permil]{overpic}
\usepackage{pict2e}

\definecolor{mycolor}{RGB}{218,232,252}
\rowcolors{2}{mycolor!60}{cyan!25}

\usepackage[subtle]{savetrees}

\begin{document}

\title{Near-Field SWIPT with gMIMO in the Upper Mid-Band: Opportunities, Challenges, and the Way Forward}

\author{\"Ozlem Tu\u{g}fe Demir, Mustafa Ozger, Ferdi Kara, Woong-Hee Lee, and Emil Bj\"ornson
\thanks{\"O. T. Demir is with Bilkent Universtity, Turkey.}
\thanks{M. Ozger is with Aalborg University, Denmark.}
\thanks{M. Ozger, F. Kara and E. Bj\"ornson are with KTH Royal Institute of Technology, Sweden. F. Kara is also with Zonguldak Bulent Ecevit University, T\"urkiye.}
\thanks{W.-H. Lee is with Dongguk University-Seoul, South Korea.}
\thanks{E. Björnson was supported by the Vinnova center SweWIN. \"O. T. Demir was supported by 2232-B International Fellowship for Early Stage Researchers Programme funded by the Scientific and Technological Research Council of T\"urkiye.}}



\maketitle

\begin{abstract}
This paper explores the integration of simultaneous wireless information and power transfer (SWIPT) with gigantic multiple-input multiple-output (gMIMO) technology operating in the upper mid-band frequency range (7–24 GHz). The near-field propagation achieved by gMIMO introduces unique opportunities for energy-efficient, high-capacity communication systems that cater to the demands of 6G wireless networks. Exploiting spherical wave propagation, near-field SWIPT with gMIMO enables precise energy and data delivery, enhancing spectral efficiency through beamfocusing and massive spatial multiplexing. This paper discusses theoretical principles, design challenges, and enabling solutions, including advanced channel estimation techniques, precoding strategies, and dynamic array configurations such as sparse and modular arrays. Through analytical insights and a case study, this paper demonstrates the feasibility of achieving optimized energy harvesting and data throughput in dense and dynamic environments. These findings contribute to advancing energy-autonomous Internet-of-Everything (IoE) deployments, smart factory networks, and other energy-autonomous applications aligned with the goals of next-generation wireless technologies.
\end{abstract}

\begin{IEEEkeywords}
 gigantic MIMO, near-field, simultaneous wireless information and power transfer, upper mid-band frequencies.
\end{IEEEkeywords}

\vspace{-4mm}
\section*{Introduction}
The 5G Radio Access Network (RAN) was designed to support high-speed mobile broadband, mission-critical communications, and massive Internet of Things (IoT) connectivity \cite{pennanen20246g}. As we move toward 6G, the number and variety of connected devices are expected to grow dramatically, signaling a shift from the IoT to the Internet of Everything (IoE). This broader vision includes sensors, machines, vehicles, drones, and robots—all connected and capable of advanced functions such as computing, sensing, positioning, and energy transfer \cite{pennanen20246g}.
In this context, improving energy efficiency and enabling battery-less, energy-autonomous devices become essential. Wireless power transfer (WPT) is seen as a key technology to power lightweight IoT/IoE networks in 6G. The first form of WPT allows devices to harvest energy from downlink signals and then use that energy to send data in the uplink. On the other hand, the most promising form of WPT is simultaneous wireless information and power transfer (SWIPT), where information and energy are transmitted together from one or more transmitters to one or more receivers. These receivers can decode information, harvest energy, or do both—either within the same device or in separate ones \cite{ashraf2021simultaneous,clerckx2018fundamentals}. SWIPT offers a balanced way to support both communication and power delivery using the same radio spectrum, paving the way for trillions of low-power, wirelessly connected devices to operate seamlessly anytime, anywhere \cite{ashraf2021simultaneous,clerckx2018fundamentals}.

Gigantic multiple-input multiple-output (gMIMO) systems, with their ability to exploit spatial diversity and increase spectral efficiency (SE) using hundreds or a thousand antennas, will likely become a cornerstone for achieving the 6G targets on capacity and energy efficiency. Operating in the upper mid-band spectrum (i.e., 7–24 GHz) offers a balance between coverage and capacity while allowing systems to harness the benefits of near-field communication, where the spherical nature of wavefront propagation can enable precise spatial focusing \cite{bjornson2024enabling}.

\begin{figure*}
		\centering
    \begin{overpic}[width=0.65\textwidth, height=0.54\textwidth]{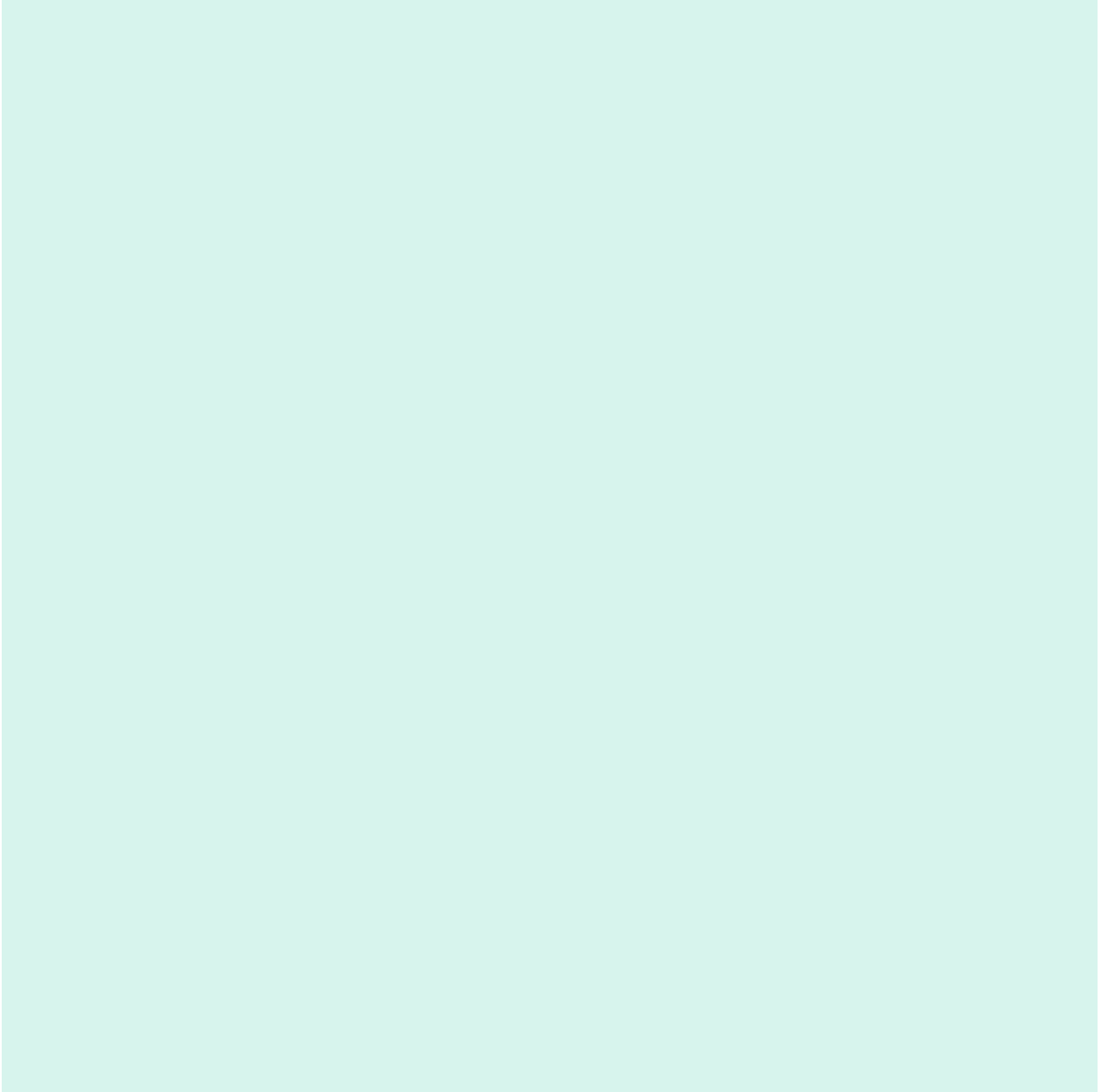}
%
\put(0,0){\includegraphics[width=0.65\textwidth]{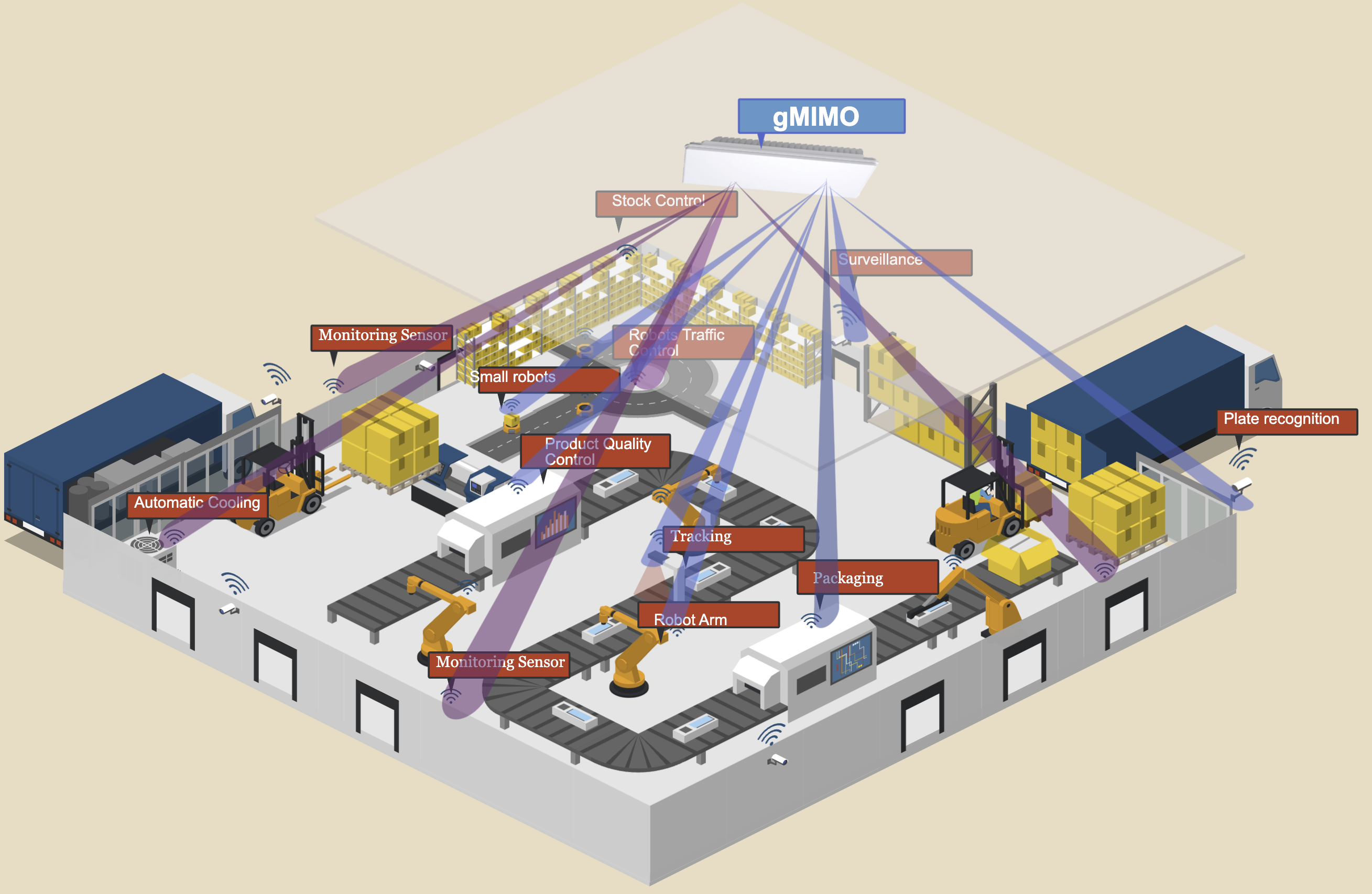}}%

\put(0,605){\includegraphics[width=11.79cm, height=3.5cm]{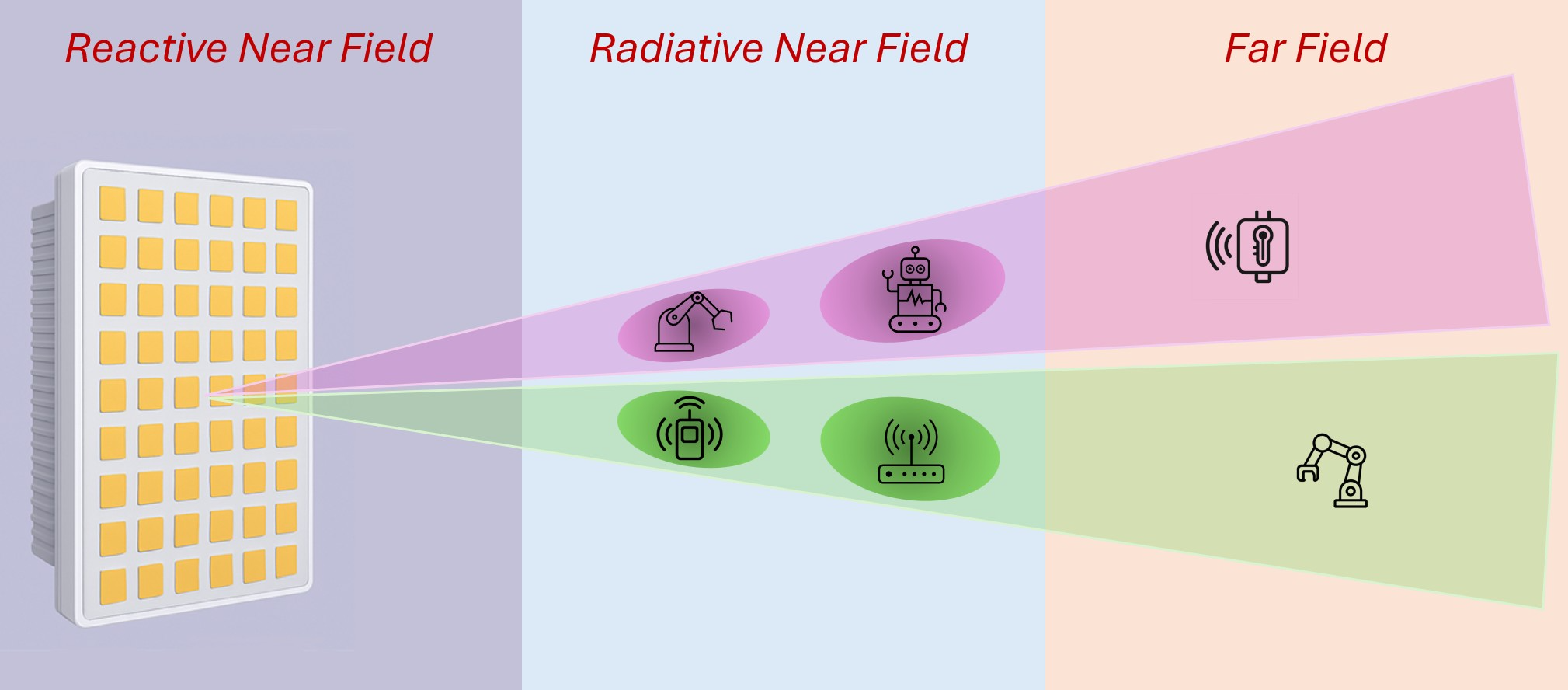}}

    \end{overpic}

    \caption{Near-field SWIPT with gMIMO in a selected use case for smart factory.}
    \label{system_model}
            \vspace{-6mm}
\end{figure*}

The integration of near-field SWIPT with gMIMO provides a unique opportunity to enhance wireless communication systems by achieving precise wireless power delivery and robust data transmission in dense environments. 
The growing adoption of smart cities, factories, autonomous vehicles, and IoT/IoE ecosystems (as depicted in Fig.~\ref{system_model}) further proves the necessity of near-field SWIPT with gMIMO. These environments demand precise energy delivery and seamless connectivity, which cannot be fully addressed by existing solutions. For instance, the integration of near-field SWIPT with gMIMO has the potential to revolutionize industrial IoT by enabling self-sustaining sensor networks in smart factories. Near-field SWIPT, leveraging the spatial focusing capabilities of gMIMO arrays, offers a compelling solution by satisfying the rate requirements of information decoding (ID) users with focused beams and allowing more transmit power to be allocated to energy harvesting (EH) devices. However, realizing this potential requires a deeper understanding of the theoretical principles, design challenges, and implementation strategies, which is the focus of this paper.

While SWIPT and MIMO have been extensively studied individually, their combination in the emerging near-field regime, particularly within the upper mid-band, remains an underexplored frontier. Most SWIPT studies focus on far-field scenarios where the plane-wave approximation simplifies analysis and implementation. These approaches often neglect the unique challenges and opportunities posed by near-field propagation. In parallel, research on near-field MIMO has primarily concentrated on enhancing spatial resolution and developing adaptive beamforming techniques, without fully examining the implications of incorporating SWIPT into these systems. This paper addresses these gaps by providing a comprehensive study of near-field SWIPT with gMIMO operating in the upper mid-band. It highlights how near-field beamfocusing and massive spatial multiplexing can be exploited to efficiently serve both information-decoding and energy-harvesting users.  Furthermore, it analyzes precoding and power allocation strategies under realistic conditions, including shadowing, imperfect CSI, and non-linear energy harvesting effects. To bridge theory and practice, the study introduces detailed application scenarios—such as smart cities, factories, and agriculture—and presents a smart factory case study demonstrating the spectral efficiency and energy harvesting gains achievable through near-field beamfocusing.

\section*{Near-field wireless propagation characteristics and power transfer}

The region around an antenna can be categorized into three zones based on the electromagnetic wave behavior: the reactive near-field, radiative near-field, and far-field. The reactive near-field, closest to the antenna, typically spans only a few wavelengths and features complex interactions between the transmitter and prospective receiver. In the radiative near-field and far-field, the radiative component is dominant. In the radiative near-field, the wave maintains a noticeable spherical shape, while in the far-field, this curvature of the waves can be approximated as planar. For an antenna array, these regions are determined by the total aperture length of the array \cite{bjornson2024towards}.

Historically, communication researchers have assumed that received electromagnetic waves arrive from the far field, where they approximate planar waves rather than spherical ones \cite{yonina}. This assumption holds because transmitter-receiver distances in practical scenarios generally exceeded the Fraunhofer distance (a.k.a.\ Rayleigh distance), which marks the boundary between the far-field and radiative near-field. The Fraunhofer distance increases with both the array’s aperture length and the operating frequency. With the advent of larger arrays and higher frequencies in modern wireless communications, radiative near-field effects are now relevant, as Fraunhofer distances can extend up to hundreds of meters \cite{bjornson2024towards}.

\vspace{-4mm}

\subsection*{Near Field versus Far Field}

One advantageous aspect of radiative near-field transmission is the ability to achieve beamfocusing, concentrating energy in both the angular and distance domains to form ellipsoidal regions around the receiver where energy is focused. In the far-field, this capability is limited to beamforming, which lacks finite-depth focusing; therefore, two users at roughly the same angle but at different distances are covered by the same beam, making it undesirable to serve them simultaneously since that would result in strong interference. Beamfocusing facilitates massive spatial multiplexing, enabling the spatial separation of users at the same angle—a feature not achievable with far-field beamforming \cite{bjornson2024towards}.

Similar to conventional communication systems, SWIPT has largely been studied in the far-field \cite{swipt_survey_2022}. Most current research on radio-frequency (RF) WPT focuses on systems operating in the sub-6 GHz band due to better propagation conditions and ``large'' isotropic receivers and utilizing a relatively small number of antennas. In practical scenarios (e.g., distances $>$ $1$–$2$ meters), charging devices are generally positioned in the far-field region \cite{7893054,near-field-wpt}. However, future 6G-based devices are expected to be located in the radiative near-field of larger arrays \cite{near-field-wpt} thanks to larger antenna aperture and higher operating frequency.
In this context, SWIPT can also leverage beamfocusing and massive spatial multiplexing. When EH users and ID users are positioned separately, near-field beamfocusing enables the system to direct signals more precisely, serving ID users with less power leakage and allowing surplus energy to be more efficiently delivered to EH users. While EH users may also benefit from harvesting ambient RF energy from information-bearing beams, the ability to spatially separate users in the near field reduces interference and allows for better beam management. In another SWIPT configuration where users perform simultaneous EH and ID, beamfocusing facilitates optimized time- or power-splitting strategies, thereby maximizing the harvested energy under joint EH-ID constraints.
\vspace{-2mm}

\section*{Near-Field SWIPT Applications and Use Cases}

SWIPT has been applied across diverse domains, particularly within the IoE \cite{lyu2022OAM}, which integrates cloud/edge computing and big data to enable advanced 6G use cases such as autonomous vehicles, drone swarms, smart cities, and smart agriculture. Beyond industrial applications, scenarios like emergency communications and wearable healthcare are also important, where battery-less devices can harvest energy from broadcast signals to maintain connectivity during disasters.

The emergence of gMIMO in the upper mid-band offers a promising approach for enabling near-field SWIPT. By accommodating a large number of antennas within a given form factor, gMIMO avoids the severe propagation challenges of millimeter-wave (mmWave) bands while benefiting from extended Fraunhofer distances that can reach tens of meters. This property allows effective near-field operation across a wide range of indoor and outdoor scenarios. For example, sensors deployed in factories, agricultural fields, and smart cities can efficiently harvest energy while maintaining reliable communication links.

\textbf{Smart cities and IoV networks:} In urban environments with autonomous vehicles and Internet of Vehicles (IoV) networks, roadside gMIMO units can simultaneously deliver high data rates and recharge sensor nodes within the network. Use cases such as traffic management, smart street lighting control, and smart public transportation can greatly benefit from the spatial precision and simultaneous energy–information transfer enabled by near-field SWIPT. Moreover, unlike far-field line-of-sight (LOS) channels, near-field propagation enables spatial multiplexing even for users at similar angles but different distances. This capability allows gMIMO arrays to support multiple vehicular or infrastructure nodes simultaneously, enhancing spectral efficiency in dense urban deployments or making more power available for EH.

\textbf{Smart factories:} In industrial environments with numerous EH users, gMIMO enables the simultaneous support of many devices within the same time–frequency resources. After meeting the communication requirements of robotic systems and controllers, surplus power can be directed to battery-less sensors, fostering a sustainable IoE ecosystem.

\textbf{Smart agriculture:} In farming scenarios, near-field SWIPT can power sensors while maintaining reliable control links with mobile ``agribots.'' Typical use cases include machinery tracking and real-time quality monitoring using distributed sensors across large fields. Moreover, since the system operates in the upper mid-band, a large number of antenna elements can be integrated into a compact gMIMO array, allowing it to be deployed at the edge of fields without occupying excessive space. This compact yet powerful array can exploit spatial diversity in both distance and angle, enhancing link reliability and enabling simultaneous support for multiple distributed devices in large-scale agricultural deployments.

In the sequel, through the extensive simulations in this paper, we will analyze a small factory environment where a gMIMO array is mounted on the ceiling as a uniform planar array, as seen in Fig.~\ref{system_model}. The system parameters are provided in Table~\ref{tab:simulation} unless otherwise stated. The EH users are equipped with a non-linear energy harvesting circuit, modeled according to \cite[Fig.~8(b)]{clerckx2018fundamentals}.

\begin{table}
\centering
\caption{Simulation Parameters.}
\begin{tabular}{|l|l|}
\hline
\multicolumn{1}{|c|}{Parameter}        & \multicolumn{1}{c|}{Value}                                                                                          \\ \hline
Carrier frequency ($f_{c}$)              & \begin{tabular}[c]{@{}l@{}}Fig. 2: $3$, $7.5$, $15$, $30$ GHz\\ Figs. 3 and 4: $7.5$, $15$ GHz\\ Fig 5: $7.5$ GHz\end{tabular} \\ \hline
Array geometries in 2D at the BS        & \begin{tabular}[c]{@{}l@{}}Fig. 2: $16$$\times$$4$, $40$$\times$$10$, $80$$\times$$20$\\ Figs. 3 and 4: $40$$\times$$10$, $80$$\times$$20$\\ Fig. 5: $40$$\times$$10$\end{tabular} \\ \hline
Antenna spacing                        & Half-a-wavelength                                                                                                   \\ \hline
Number of ID users                     & \begin{tabular}[c]{@{}l@{}}Fig. 2: $\{10, 20, 30, 40\}$\\ Fig. 3: $\{5, 10, 15, 20\}$\\ Fig. 4: $20$\\ Fig. 5: $1$\end{tabular}          \\ \hline
Number of EH users                     & \begin{tabular}[c]{@{}l@{}}Figs. 2, 4, and 5: $10$\\ Fig. 3: $5$\end{tabular}                                           \\ \hline
Transmit power of the base station     & $10$ W                                                                                                                \\ \hline
Bandwidth                              & $100$ MHz                                                                                                             \\ \hline
Noise power spectral density           & $-204$ dBW/Hz                                                                                                          \\ \hline
Noise figure &  $7$ dB
         \\ \hline
Uplink pilot power                     & $100$ $\mu$W                                                                                                           \\ \hline
Number of symbols in a coherence block & $10\,000$                                                                                                                 \\ \hline
Analog circuit power of EH devices & $3$\,mW \\ \hline
\end{tabular} \label{tab:simulation}
\vspace{-2mm}
\end{table}

\section*{Opportunities}

\subsection*{Gigantic MIMO in Upper Mid-band}

The initial 5G deployments operate at a frequency of $3.5$ GHz. However, frequencies in the upper mid-band, such as $6.425$-$7.125$ GHz, have also been identified as potential candidates for next-generation wireless systems. To meet the ever-growing data rate demands of applications like immersive reality and holographic communications, researchers are exploring even higher frequencies \cite{yonina}. mmWaves spanning $24$ GHz and above are considered a natural extension due to the abundance of available spectrum in these ranges. Consequently, 5G research primarily focused on two frequency ranges: FR1 ($410$-$7125$ MHz) and FR2 ($24.25$–$71.0$ GHz) \cite{ITU2023}.  

While FR1 is already heavily utilized and crowded, FR2 faces significant practical challenges, including high susceptibility to blockages and inefficient RF hardware. An alternative is emerging in the form of the upper mid-band frequencies ($7.125$–$24$ GHz), also known as FR3. The upper mid-band offers an attractive compromise, providing more bandwidth than FR1 and better propagation characteristics than FR2. Moreover, this range supports a greater number of antenna elements without altering the physical size of the whole array, still resulting in near-field effect advantages for spatial multiplexing and beamfocusing. Additionally, employing multiple antennas at the receiver side can enhance the spatial degrees-of-freedom \cite{bjornson2024enabling}.

\begin{figure}[t!]
	\hspace{-2cm}
		\begin{center}
			\includegraphics[trim={10mm 2mm 18mm 8mm},clip,width=3.4in]{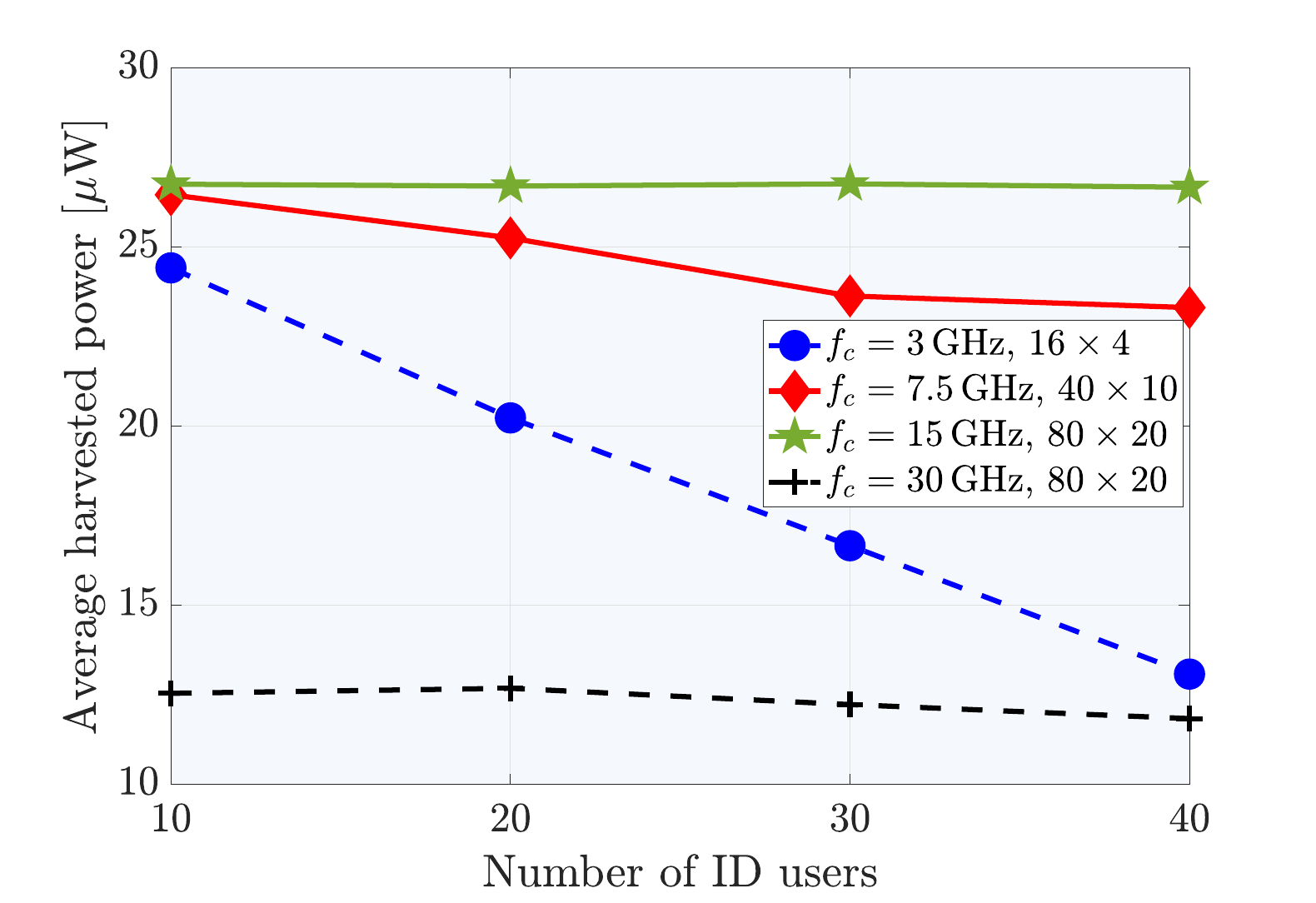}
            \vspace{-2mm}
			\caption{Average harvested power versus the number of ID users for different bands.} \label{fig:fig2}
		\end{center}
                \vspace{-8mm}
\end{figure}

Upper mid-band frequencies can meet both the performance goals of 6G and the unique demands of SWIPT systems by balancing energy harvesting and communication. Spatial multiplexing improves SE for ID users, while wireless power transfer primarily benefits from high beamforming gain and near-field focusing, which enable precise and efficient energy delivery to EH users. To this end, we analyze the EH capability of a SWIPT system under various configurations, including different frequency bands, numbers of ID users, and antenna configurations at the serving base station (BS) in Fig.~\ref{fig:fig2}. The path loss model for an indoor factory environment from \cite{3gpp2018study} is adopted, incorporating lognormal shadowing with a standard deviation of \(4\,\text{ dB}\). The gMIMO array is strategically positioned on the ceiling to ensure an LOS connection to each user.
 Correlated Rician fading with isotropic scattering is considered, where the \(\kappa\)-factor follows a lognormal distribution with a mean of \(7\,\text{ dB}\) and a standard deviation of \(8\,\text{ dB}\). 10 EH users are randomly deployed, while the horizontal axis in Fig.~\ref{fig:fig2} denotes the number of ID users served simultaneously. 
The distances, azimuth angles, and elevation angles of all EH and ID users are randomly generated across multiple realizations. The distances and angles are sampled uniformly within the range $[1.5-25]$ meters and  $[-\pi/2,\pi/2]$, respectively.
The BS has a total power budget of $10$\,W, with antennas arranged in a uniform planar array as indicated in the legend. The inter-antenna spacing is set to half a wavelength. The system operates over a bandwidth of \(100\,\mathrm{ MHz}\), and zero-forcing precoding is employed under imperfect channel state information (CSI), where the least-squares (LS) estimator is used.

The performance is evaluated in terms of spectral efficiency (SE) for ID users and average harvested power for EH users. We examine different array configurations, operating frequencies (as summarized in Table \ref{tab:simulation}), and channel models (near-field vs. far-field) to assess their impact on energy transfer and spatial multiplexing capabilities. The specific parameter settings corresponding to each figure are listed in Table \ref{tab:simulation}.

A power allocation scheme ensures a minimum SE of $4$\,bit/s/Hz for each ID user, with the remaining power budget used to transmit energy beams. These energy beams are designed to lie in the null space of the ID user channel estimates to reduce interference. Although the current design employs separate beams for data and energy transmission, future research could investigate the joint optimization of information-carrying signals to simultaneously support EH, thus reducing the reliance on dedicated energy beams. 

Fig.~\ref{fig:fig2} illustrates four configurations, with the first three maintaining the same physical array area but operating at increasing frequencies, thereby accommodating a higher number of antennas. The upper mid-band frequencies with a gMIMO array show significantly improved harvested power per EH user, particularly as the number of ID users increases. This is evident in the energy saved through more focused data transmission to ID users, allowing that energy to be used to serve more EH users. This gain stems from the fact that higher frequencies allow packing more antennas into the same physical array area, which results in narrower beams and higher array gain.  
Consequently, achieving the SE targets for ID users requires less transmit power, allowing a greater portion of the power budget to be allocated to energy transmission. In contrast, for mmWave bands (i.e., the case when $f_{c}=30$ GHz in Fig.~\ref{fig:fig2}), the number of antennas is capped at $80 \times 20 = 1600$ due to hardware limitations. The reduced form factor at these higher frequencies restricts the advantages of adding more antennas, which diminishes the efficiency gains compared to the upper mid-band. 
While larger mmWave arrays could in theory achieve even higher gains, such arrays are less feasible in practice due to increased cost and complexity. Hence, gMIMO in the upper mid-band emerges as a promising solution, offering an optimal balance between hardware complexity and enhanced spatial multiplexing capabilities.

\vspace{-2mm}

\subsection*{Beamfocusing and Massive Multiplexing for SWIPT}

The near field presents numerous opportunities, with one of the most significant being the capability for beamfocusing. This allows transmission to be directed toward specific locations, enabling beam formation in both the angular and distance domains, akin of how a lens focuses light. For instance, even if two receivers are positioned along the same angular line, their differing distances from the transmitter prevent interference between their communications. This characteristic offers a unique advantage for designing spatial beamforming strategies. 

\begin{figure}[t!]
	\hspace{-2cm}
		\begin{center}
			\includegraphics[trim={10mm 2mm 18mm 8mm},clip,width=3.4in]{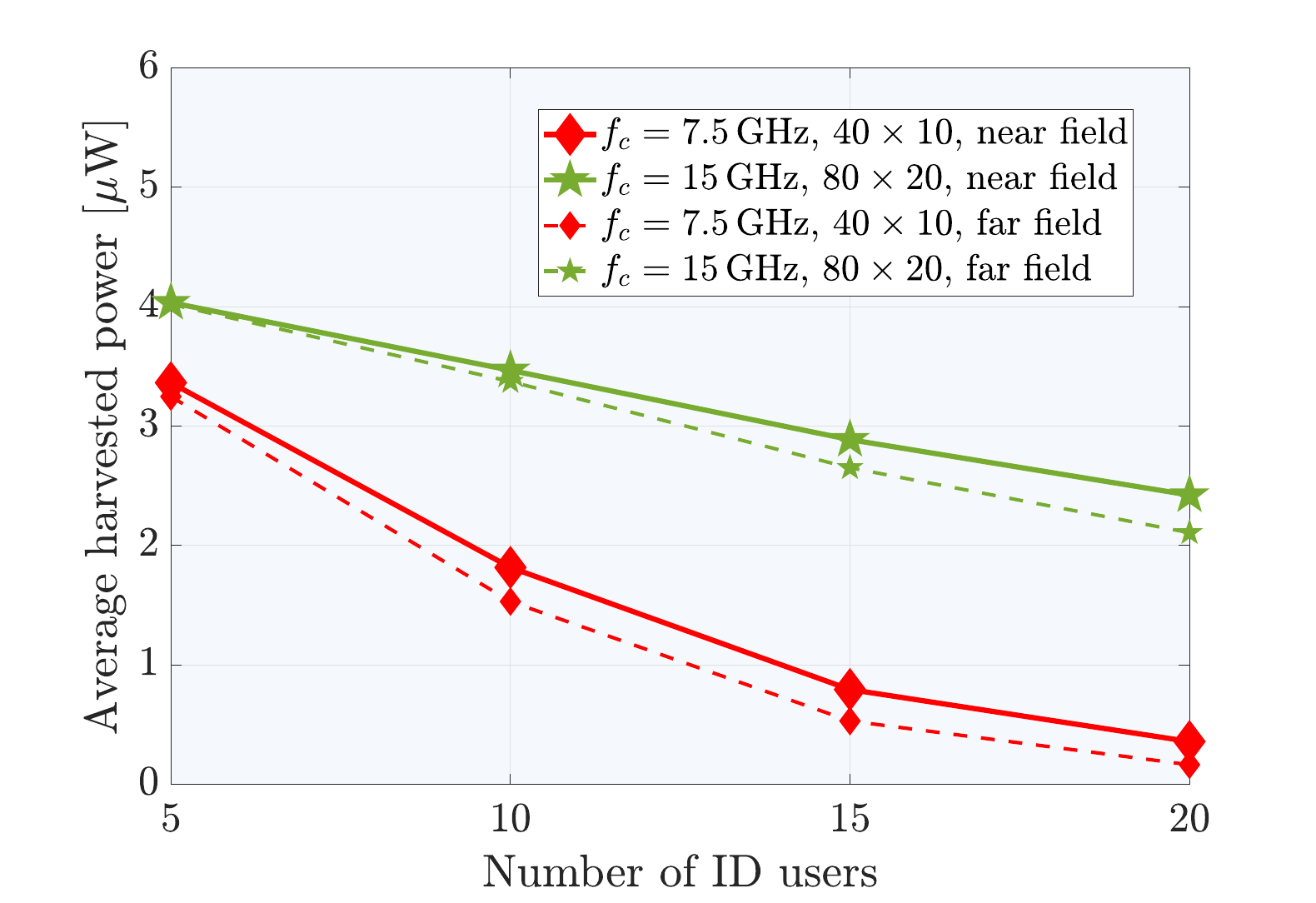}
            \vspace{-2mm}
			\caption{Received power versus the number of ID users for near- and far-field channels.} \label{fig:fig3}
		\end{center}
                \vspace{-8mm}
\end{figure}

To illustrate how near-field beamfocusing enhances channel properties and increases EH capability, Fig.~\ref{fig:fig3} compares the far-field and near-field scenarios under perfect CSI assumption. For a fair comparison of the channel structure, all channel vectors are scaled to correspond to a path loss at a distance of $25$\,m. Thus, the observed differences in Fig.~\ref{fig:fig3} arise solely from the planar and spherical wave characteristics inherent to the far-field and near-field regions, respectively. All users are positioned within a narrow region near the broadside direction of the gMIMO array to highlight the advantages of near-field massive spatial multiplexing. As shown in Fig.~\ref{fig:fig3}, under identical path loss conditions, near-field propagation enables greater average harvested energy. This improvement stems from the more efficient utilization of transmit power and the increased spatial multiplexing capability offered by near-field focusing. With more focused beams, the system can more easily satisfy the SE requirements of ID users using less power, which leaves more power available for EH.

\vspace{-2mm}

\section*{Challenges}

\subsection*{Near-field Gigantic MIMO Channel Estimation}

For the effective and efficient operation of SWIPT, leveraging the spatial multiplexing and beamforming gains of gMIMO, it is crucial to estimate the channels of both ID and EH users. Channel estimation in near-field gMIMO systems is challenging due to the unique characteristics of these systems. One primary challenge arises from the spherical wavefront in the radiative near-field region, which complicates channel modeling and estimation. While reciprocity-based beamforming can be applied without explicitly modeling near-field characteristics, many estimation techniques rely on parameterized models that assume planar wavefronts. These models are no longer valid in the near field, where channel sparsity shifts from the angular to the polar domain, incorporating both distance and angular components \cite{Cui2023near}. While polar domain representation can improve sparsity modeling, it introduces complexity due to the increased number of parameters.

Accurate acquisition of CSI is critical for directing beams toward users. However, the increased number of users and user-side antennas introduces substantial pilot overhead. Another significant challenge is the non-stationarity of the channels. With a large number of antenna elements, spatially separated elements experience distinct channels characterized by different multipath profiles. Current near-field channel estimation algorithms often assume spatial stationarity, where all antenna elements observe the same scatterers. In reality, different regions of the array may experience distinct scatterers, limiting the effectiveness of these algorithms and reducing the quality of channel estimation.  
To address these challenges, recent research has focused on dividing the antenna array into spatially stationary sub-arrays (i.e., modular arrays), allowing channel estimation algorithms to extract overall channel information \cite{modular_array}. 

\vspace{-2mm}

\subsection*{Design of SWIPT Precoding and Power Allocation}

To meet the data rate and energy requirements of ID and EH users, respectively, optimized precoding and power allocation are essential. 

In a gMIMO system operating in the upper mid-band, BSs can leverage a massive number of antennas, while user terminals can also be equipped with multiple antennas in a compact form factor. This evolution introduces unique challenges and opportunities for designing SWIPT precoding and power allocation. Conventional algorithms tailored for single-antenna users must be reconsidered for multiple-antenna scenarios, necessitating the development of innovative precoding and power allocation strategies.

For systems using the power splitting protocol, it is crucial to develop adaptive algorithms that dynamically adjust the power splitting ratio in real time based on battery levels, channel conditions, data requirements, and environmental factors. The SE level can be adaptively adjusted based on users’ remaining energy and predicted traffic patterns. By employing reinforcement learning (RL), the system can anticipate future packet activity and dynamically optimize the power-splitting ratio and precoding strategy. For example, in a smart factory, user activity often follows predictable patterns, which can be effectively learned by RL agents to enable proactive resource adaptation.

Future systems may incorporate hybrid power splitting methods that combine orthogonal power splitting for EH (WPT) and non-orthogonal power splitting for data transmission (SWIPT). Such approaches can enable a more efficient trade-off in scenarios with fluctuating data demand.

\vspace{-2mm}

\subsection*{Hardware Cost}

The hardware costs of a gMIMO SWIPT system can be analyzed from both the BS and user terminal perspectives. At the BS, the presence of a gigantic number of antennas increases hardware complexity and processing costs compared to conventional massive MIMO systems. Fully digital beamforming is envisioned as the preferred approach to fully exploit the near-field beamfocusing and massive spatial multiplexing capabilities of gMIMO. However, cost-efficient hybrid beamforming architectures, including recently proposed tri-hybrid structures, can also be explored to reduce the number of RF chains while aiming to preserve most of the near-field processing gain. Designing such architectures to balance hardware cost and near-field performance constitutes an important research question for future work.
On the user side, ID user terminals equipped with a greater number of antennas also experience increased hardware costs. Similarly, EH users can benefit from multiple rectennas—antennas that rectify signals— which enhance harvested energy but bring additional hardware costs.

Enhancing the efficiency of EH systems is crucial for the effective deployment of SWIPT, particularly in close-proximity settings. This can be achieved by developing advanced materials for rectennas capable of capturing energy over a broader spectrum and improving the conversion of electromagnetic energy into usable direct current power. Additionally, exploring multi-modal EH techniques that utilize multiple frequencies for both power and data transmission could significantly enhance system efficiency. Such approaches are especially valuable in densely populated urban environments, where spatial diversity plays a critical role. However, manufacturing this sophisticated hardware is still in the early stages and might be costly for mass production. 

\vspace{-2mm}

\section*{Enabling Solutions and Future Research Directions}

\subsection*{Near-field-aware Low-pilot Channel Estimator Design}

As discussed earlier, near-field propagation introduces additional modeling complexity, particularly for parametric estimators, due to spherical wavefronts and the spatial diversity in gMIMO systems. Moreover, challenges such as reduced coherence time and lower SNR per antenna in the estimation phase affect both parametric and classical estimators. Advancing channel modeling and estimation techniques with a reasonable allocation of pilot resources is crucial for improving system performance.
Developing realistic and scalable near-field channel models that accurately capture spherical wave propagation and interactions between the transmitter, receiver, and surrounding environment is essential. These models should account for reflection, diffraction, and scattering effects to simulate a wide range of real-world conditions effectively.

The upper mid-band frequencies exhibit sparser propagation characteristics compared to sub-6\,GHz systems, which simplifies channel estimation using parametric techniques. In the considered near-field channel model, each user’s LOS channel component can be described by three unknown parameters: one distance and two angular variables. As a result, the number of required pilot symbols is on the same order as the number of unknown parameters—typically only a few symbols per user—rather than scaling with the number of antennas as in conventional LS estimation. This leads to a substantial reduction in pilot overhead for large gMIMO arrays. Classical direction-of-arrival algorithms and compressed sensing-based techniques can be effectively employed to estimate these few channel parameters.

\begin{figure}[t!]
	\hspace{-2cm}
		\begin{center}
			\includegraphics[trim={6mm 0mm 18mm 8mm},clip,width=3.4in]{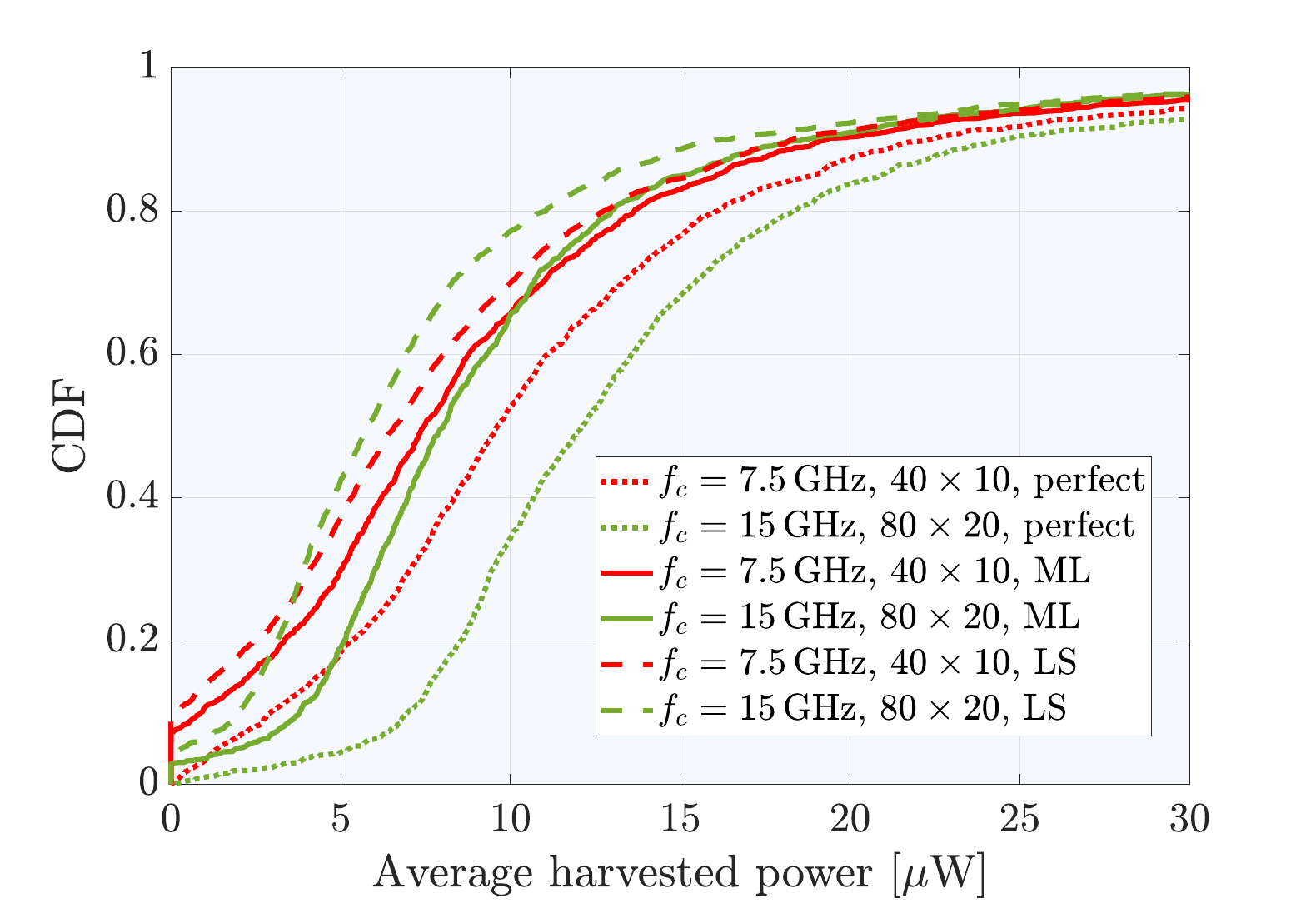}
            \vspace{-2mm}
			\caption{The CDF of average harvested power with perfect CSI and different channel estimators.} \label{fig:fig4}
		\end{center}
        \vspace{-8mm}
\end{figure}

To observe the effect of channel estimation errors on the harvested energy, we consider a gMIMO system with $20$ ID users and $5$ EH users in Fig.~\ref{fig:fig4}. Under the requirement for SE of $4$\,bit/s/Hz for the ID users, the cumulative distribution function (CDF) of the net average harvested power per EH user is plotted. Initially, the EH users transmit orthogonal pilot sequences of $5$ symbols to the gMIMO BS, which then estimates their channels using either a parametric maximum likelihood (ML)-based channel estimator or a least-squares (LS) channel estimator. The analog circuit power consumption of each EH device is set to $3$\,mW, consistent with the low-power IoT device specifications in \cite{morteza}.
As shown in Fig.~\ref{fig:fig4}, the CDF curves corresponding to the ML estimator are shifted further to the right, indicating that more power can be harvested when using the ML estimator, as it provides higher channel estimation accuracy. However, this improvement comes at a significant computational cost due to the three-dimensional grid search required in ML estimation, which makes the number of operations scale cubically with the grid resolution. In contrast, the LS estimator involves only a straightforward matrix multiplication. Striking a balance between estimation accuracy and computational complexity remains an important research direction.

\vspace{-2mm}
\subsection*{Joint Precoding of Information and Dedicated Energy Symbols}

Energy beamforming, similar to data beamforming, is a promising area that requires further exploration. By directing energy beams toward specific receivers, systems can efficiently deliver power to low-energy devices while maintaining high data throughput for ID users. Future research should focus on optimizing energy beamforming in complex near-field scenarios to maximize efficiency and performance.

Joint precoding of information and dedicated energy symbols can be optimized based on CSI. In some cases, dedicated energy symbols may not be necessary, as EH users can meet their EH requirements from existing or redesigned information beams to also be partially captured by EH receivers. However, these redesigned information beams may cause a sacrifice in the quality-of-service requirements of the ID user. Such design considerations may lead to different conclusions in near-field SWIPT scenarios, where beamfocusing can be employed as an alternative to far-field beamforming.

Optimizing precoding for multiple-antenna ID and EH users presents another compelling research direction. Modified versions of the waterfilling power allocation could be developed to maximize the objectives of SWIPT systems. The choice of optimization metric will significantly influence the resulting solutions. For instance, EH users need to send pilot signals to enable the gMIMO array to accurately estimate channels and beamfocus on EH users. However, transmitting pilots consumes additional energy, making the design of pilot length and power a critical optimization variable in joint precoding strategies.

\begin{figure}[t!]
    \centering
        \includegraphics[trim={6mm 2mm 14mm 8mm},clip,width=\columnwidth]{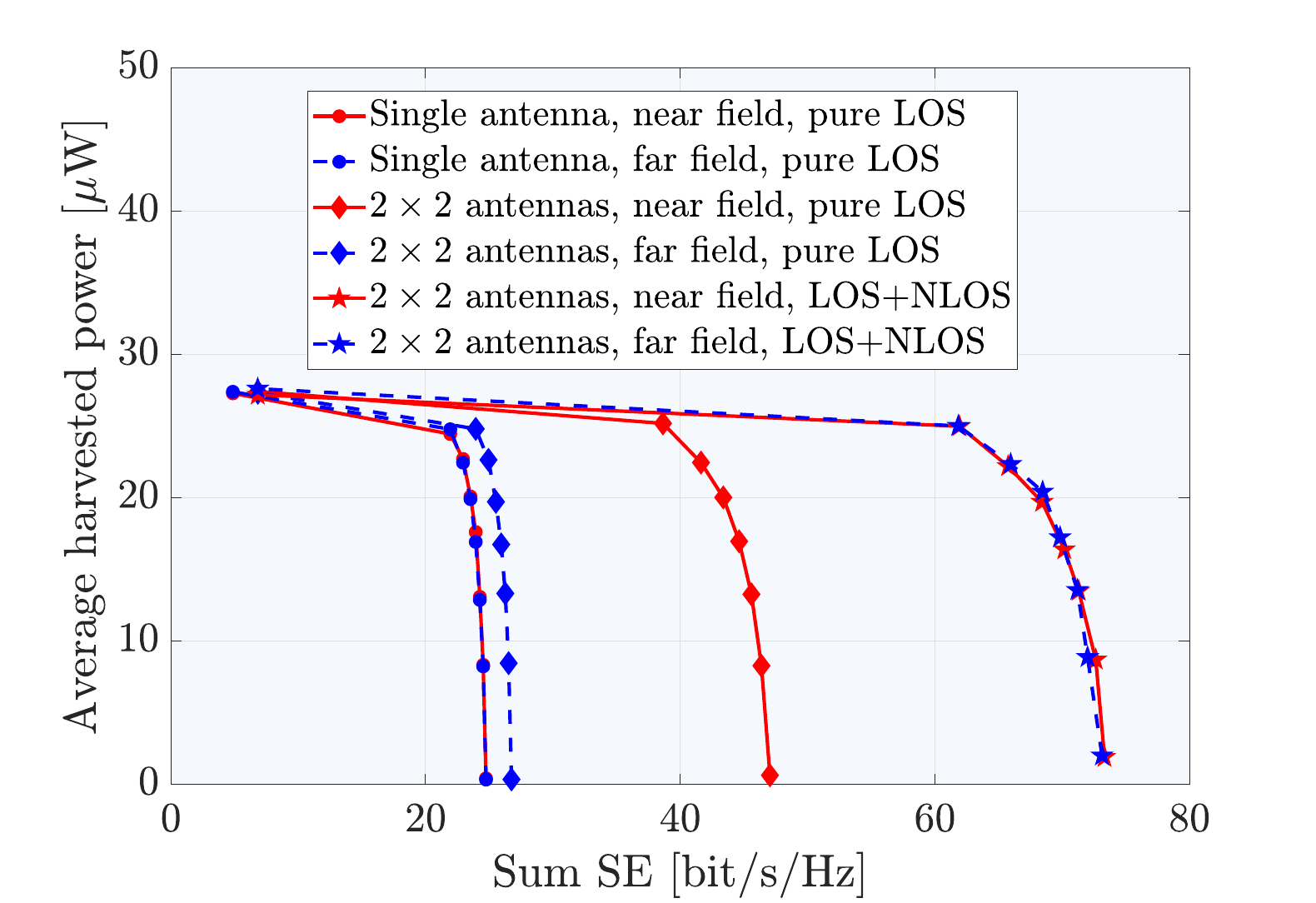}
        \includegraphics[trim={6mm 2mm 14mm 8mm},clip,width=0.48\textwidth]{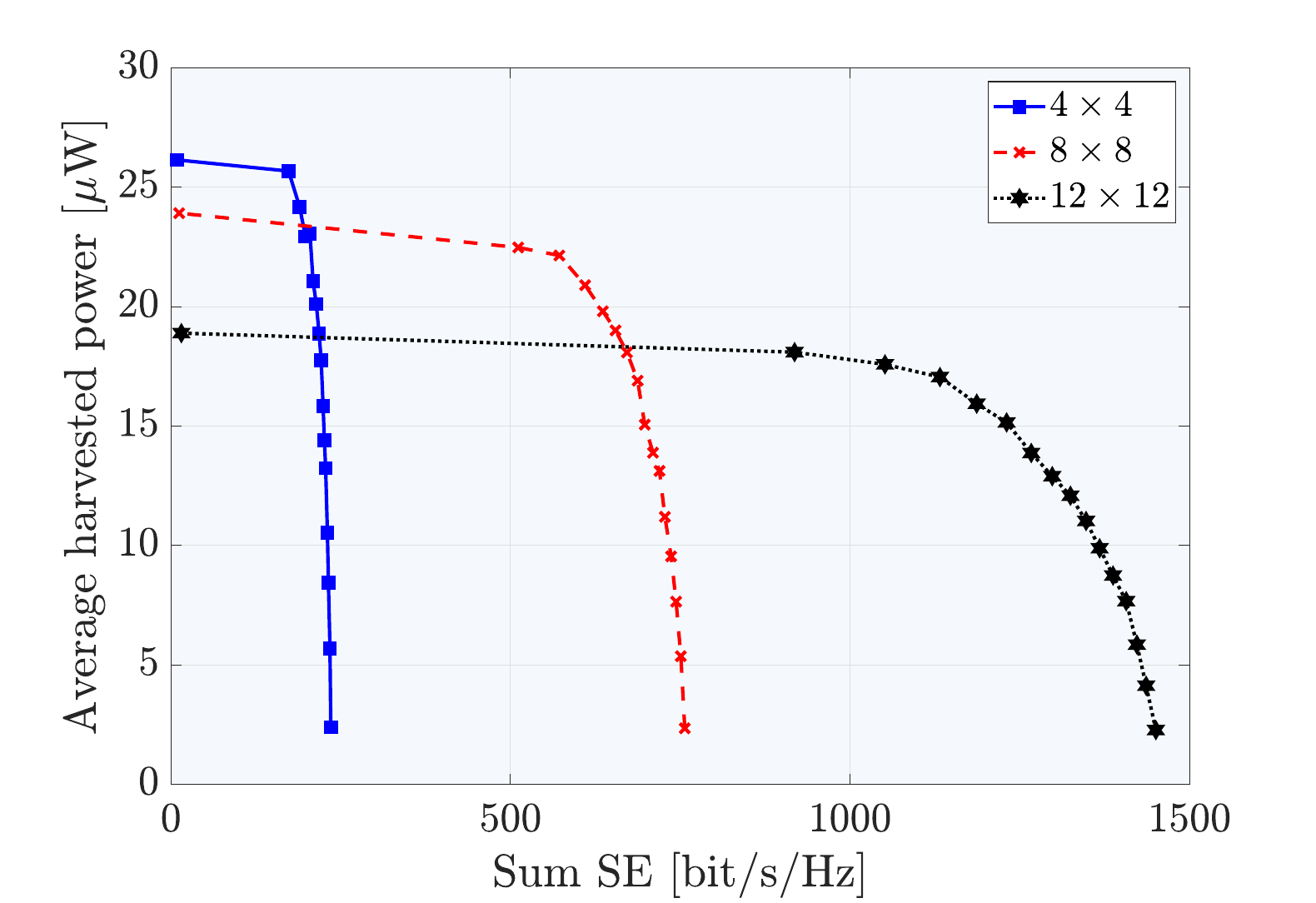}
        \caption{Average harvested power vs. sum SE for a small factory environment with different cases.}
    \label{fig:fig_56}
            \vspace{-8mm}
\end{figure}

Focusing on our case study, i.e., the smart factory use case, we investigate the average harvested power versus the SE achieved by the ID user as seen in Fig.~\ref{fig:fig_56}. The gMIMO array provides data to a robotic arm located directly below the array at a distance of $25$\,m. The robotic arm serves as the sole ID user, while $10$ sensor nodes, functioning as EH users, are randomly distributed within the $[1.5, 25]$\,m vicinity of the array. These curves are generated by varying the power allocation between the ID user and the EH users, with a water-filling power allocation applied to the ID MIMO system.

In the first scenario in the top part of Fig.~\ref{fig:fig_56}, the ID user has a single antenna, and we evaluate both near-field and far-field array response vectors under a pure LOS condition. For a fair comparison, the far-field case assumes the same path loss as if the ID user were at a distance of $25$\,m. As the figure indicates, there is no significant difference in the harvested power-SE trade-off when the ID user has a single antenna. However, when the number of antennas at the ID user is increased, a substantially higher sum SE can be achieved through the multiple antennas of the ID user for a given EH requirement.
In the pure LOS scenario, the near-field response shows a clear advantage over the far-field response. This is because a near-field LOS scenario can achieve a high-rank MIMO channel, enabling spatial multiplexing, whereas a far-field MIMO channel is limited to at most rank one, which constrains spatial multiplexing. When additional non-LOS (NLOS) paths with isotropic scattering are introduced, the gap between near-field and far-field performance diminishes, and a significantly higher sum SE is observed. This result holds even under a high $\kappa$-factor of $20$\,dB. However, it should be noted that scenarios with higher spatial correlation than isotropic scattering may yield different outcomes.

In the lower part of Fig.~\ref{fig:fig_56}, we consider near-field SWIPT with a higher number of ID user antennas, as indicated in the legend. Interestingly, increasing the number of ID user antennas decreases the maximum average harvested power showing a specific trade-off relationship in the near-field multi antenna SWIPT scenarios. This can be attributed to the fact that the energy beamformers are designed in the null space of the ID precoding matrix, which has a higher dimension as the number of subchannels increases in a near-field scenario. Consequently, the dimensionality of the space available for the energy precoder becomes more restricted when the number of antennas at the ID user increases.
This observation is noteworthy, as the optimal number of antennas may vary depending on the SE requirement of the robotic arm. For lower SE requirements, fewer antennas can be activated, which leads to a modified version of the classical water-filling algorithm. This adaptation makes optimization for multi-antenna users an intriguing research direction.

\vspace{-2mm}
\subsection*{Sparse Array and Modular Array Design}
Sparse arrays have been extensively studied in the field of array signal processing. In a sparse array, sensors or antennas are arranged in a (non)uniform manner, with part of/all elements spaced more than half a wavelength apart. This design allows for an increased array aperture with a fixed number of antennas, improving the array resolution. Sparse arrays also offer a potential solution to the hardware cost challenges of gMIMO systems, where the goal is to deploy antennas within a given array aperture to preserve the near-field effects for SWIPT capabilities of the gMIMO array. By adopting sparse array designs, the cost can be reduced compared to dense antenna placement within the array aperture. Furthermore, sparse arrays can be dynamically configured by selecting antennas from a dense antenna grid based on real-time CSI and the locations of EH and ID users. In industrial environments, ceiling-mounted sparse arrays may be effectively realized by selecting an appropriate subset of elements from a dense grid, where the choice is guided by real-time CSI and the spatial distribution of EH and ID users, thereby enabling dynamic adaptation to the prevailing network conditions.

As previously mentioned, modular arrays (also referred to as modular MIMO) use groups of densely packed antennas arranged sparsely across the aperture. This structure is particularly suitable for distributed deployments, such as modules mounted along a building facade or on factory pillars. Beyond hardware cost reduction, modular arrays enable the development of novel distributed SWIPT precoding algorithms by leveraging cost-efficient local processing at each subarray. Future research should also explore advanced rectennas with higher RF-to-DC conversion efficiency to further reduce energy consumption while maintaining performance.

\vspace{-2mm}

\section*{Conclusion}

The integration of near-field SWIPT with gMIMO in the upper mid-band presents a transformative pathway for achieving the energy efficiency and high data capacity required by 6G and beyond. By leveraging beamfocusing and massive spatial multiplexing, gMIMO systems can enable simultaneous energy harvesting and data transmission with unparalleled precision. These gains stem from the ability to pack more antennas into a fixed aperture at higher frequencies, enabling narrower beams and higher array gain. This allows ID users to be served more efficiently, leaving more power for energy harvesting. Additionally, near-field beamfocusing improves spatial separation, making it easier to meet ID users’ quality-of-service requirements while supporting energy delivery. This paper highlights key challenges, including the need for near-field-specific channel estimation, adaptive precoding techniques, and cost-effective array designs. Solutions such as sparse and modular arrays, as well as joint precoding of information and dedicated energy symbols, are proposed to address these challenges and enhance system performance. The case study focusing on smart factories illustrates how gMIMO arrays can be deployed in practical scenarios, offering valuable insights into optimizing SWIPT systems. Future research should focus on mobile scenarios, distributed processing, multi-modal energy harvesting, and scalable implementations with experimental validation, and large-scale testbed deployment to fully unlock the potential of this promising technology, paving the way for sustainable and efficient wireless communication networks.

\bibliographystyle{IEEEtran}
\bibliography{IEEEabrv.bib,references.bib}
\end{document}